\documentclass[manuscript]{aastex61}

\newcommand\adspr{AdSpR}
\newcommand\ssrv{SSRv}
\newcommand\angeo{AnGeo}

\usepackage{graphicx}
\usepackage{amssymb}
\usepackage{natbib}

\received{July 1, 2016}
\revised{September 27, 2016}
\accepted{\today}
\submitjournal{ApJL}

\shorttitle{Ion populations at an IP shock}
\shortauthors{Kajdi\v{c} et al.}

\begin{document}

\title{Different types of ion populations upstream of the 8 October 2013 interplanetary shock}

\correspondingauthor{Primo\v{z} Kajdi\v{c}}
\email{primoz@geofisica.unam.mx}

\author[0000-0002-0625-8892]{Primo\v{z} Kajdi\v{c}}
\affil{Instituto de Geofi\' isica, Universidad Nacional Aut\'onoma de M\'exico, Circuito de la investigaci\'on Cient\' ifica s/n, Ciudad Universitaria, Delegaci\'on Coyoac\'an, C.P. 04510, Mexico City, Mexico}

\author[0000-0002-3039-1255]{Heli Hietala}
\affil{Department of Earth, Planetary, and Space Sciences, University of California, Los Angeles, California, USA}

\author{X\'ochitl Blanco-Cano}
\affil{Instituto de Geofi\' isica, Universidad Nacional Aut\'onoma de M\'exico, Circuito de la investigaci\'on Cient\' ifica s/n, Ciudad Universitaria, Delegaci\'on Coyoac\'an, C.P. 04510, Mexico City, Mexico}

\begin{abstract}
We show for the first time that different types of suprathermal ion distributions may exist upstream of a single interplanetary shock. ACE and the two ARTEMIS satellites observed a shock on 8 October 2013. The ARTEMIS P1 and P2 spacecraft first observed field-aligned ions (P1) and gyrating ions (P2) arriving from the shock. These were followed by intermediate ions and later by a diffuse population. At the location of the P2 the shock exhibited an Alfv\'enic Mach number of M$_A$=5.7 and was marginally quasi-perpendicular, ($\theta_{Bn}$=47$^\circ$). At P1 spacecraft the shock was weaker (M$_A$=4.9) and more perpendicular ($\theta_{Bn}$=61$^\circ$). Consequently the observed suprathermal ion and ultra low frequency wave properties were somewhat different. At P2 the ULF waves are more intense and extend farther upstream from the shock. 
The energies of field aligned and gyrating ions in the shock rest frame were $\sim$20~keV, which is much more than in the case of the stronger (M$_A$=6-7) Earth's bow-shock, where they are less than 10~keV.
\end{abstract}

\keywords{solar wind --- interplanetary medium --- shock waves --- acceleration of particles --- waves}

\section{Introduction} \label{sec:intro}

High energy particles, such as Solar Energetic Particles \citep[SEP; e.g., ][]{schwenn06, reames96} and Energetic Storm Particles \citep[ESP; e.g., ][]{cohen06}, are common in the Solar System. It is important to study them since they present hazard for spacecraft, humans in space and even our ground based technologies such as power grids. The SEPs are also interesting since they can be used to study elemental and isotopic composition of the Sun and particle acceleration mechanisms \citep{williams98}.

Significant accelerators of energetic particles in the Solar System are collissionless shocks which belong to two major groups: planetary and interplanetary (IP) shocks. Planetary shocks form when the solar wind (SW) encounters obstacles such as planets with intrinsic magnetospheres \citep[e.g., Mercury, Earth, Saturn, Jupiter; see for example, ][]{bagenal92, russell93}, planets with induced magnetospheres such as Venus and Mars \citep[e.g., ][]{luhmann04}
and active comets \citep[e.g., ][]{cravens04}. Due to their form, planetary shocks are also called bow-shocks. The major drivers of the IP shocks are interplanetary coronal mass ejections \citep[ICME; ][]{sheeley85} and stream interaction regions \citep[SIR; ][]{gosling99}. Especially the ICME driven IP shocks have been recognized as important accelerators of energetic particles \citep[e.g., ][]{kahler03, manchester05}.

When the fast magnetosonic Mach number M$_{ms}$ of a collisionless shock exceeds a certain critical value M$_c$, the shock is called supercritical. The M$_c$ depends on several parameters, such as the angle between the direction of the upstream interplanetary magnetic field (IMF) and the shock normal, $\theta_{Bn}$ \citep{edminston86}. The supercritical shocks dissipate the kinetic energy of the incoming SW by energizing and reflecting a portion of the incident particles (ions, electrons) back upstream. Shocks are further divided according to $\theta_{Bn}$. For $\theta_{Bn}<$45$^\circ$ ($\geq$45$^\circ$), they are called quasi-parallel (quasi-perpendicular). In the case of the Earth's bow-shock the reflected ions have been observed for $\Theta_{Bn}\leq$70$^\circ$ \citep[e.g., ][]{eastwood05}. These are also called backstreaming particles. Interaction of backstreaming ions with the incident SW ions results in the growth of ultra-low frequency (ULF) waves \citep[e.g., ][]{dorfman17}. At Earth these waves have periods of $\sim$30~s on average. The region upstream of quasi-parallel shocks populated with ULF waves (suprathermal ions) is called the ULF wave (suprathermal ion) foreshock \citep[e.g., ][and references therein]{eastwood05}.

In the case of Earth there are plenty of observations of backstreaming particles. Near the leading edge of its foreshock a spacecraft first observes {\it field-aligned ion beams} \citep[FAB; ][]{gosling78, gosling79, thomsen85, kis07, meziane13}. These ions stream upstream along the IMF and exhibit highly collimated, beam-like distributions in velocity space. Their energies are below 10~keV and they are not acompanied by ULF waves although they are responsible for their generation \citep{thomsen85, eastwood05}. The FABs are also considered to be the seeds of the so called {\it diffuse ions} \citep[e.g., ][]{fuselier86, kis04}, which show almost isotropic distributions in the SW frame with a small average bulk velocity directed sunward. These ions are observed upstream of the almost parallel section of the Earth's bow-shock, they exhibit energies up to several hundreds of keV, and are accompanied by compressive ULF fluctuations. The third kind of suprathermal ions is called {\it intermediate} \citep[][]{paschmann79} with distributions intermediate between the FABs and diffuse ions. They are thought to form because farther from the edge of the foreshock the ULF waves disrupt the FAB ions, scattering them in pitch angle (PA) which leads to crescent-shaped and later to diffuse distributions. Other ion distributions have also been observed: \citet{paschmann82} observed the so called {\it gyrating ions} that exhibit distribution peaks at non zero PAs relative to the IMF. Special cases of gyrating distributions are {\it gyrotropic ions} with distribution being a torus with a symmetry axis parallel to the IMF direction \citep{winske84} and {\it gyrophase-bunched ions} \citep{gurgiolo81, gurgiolo83, eastman81, thomsen85}.

In order to distinguish between the FABs and the gyrating ions we use criteria similar to \citet[][]{savoini13} and references therein. Backstreaming ions are classified as FABs if they exhibit picth angles between $\sim$0$^\circ$ and $\sim$30$^\circ$, while they are denominated as gyrating ions if their pitch angles extend to larger values (e.g., $\sim$90$^\circ$).

Although ions in suprathermal particle energy range have been observed upstream of IP shocks at 1~A.U., they mostly exhibit diffuse distributions \citep[e.g., ][]{armstrong70, bavassano86, gosling83, gosling84}. It is not clear whether these ions were actually accelerated by IP shocks near 1~AU or whether they are just low-energy parts of SEPs.

Only two works report observations of ion populations other than diffuse upstream of IP shocks: \citet{vinas84} show ion spectra upstream of an IP shock observed on 3 February 1978 obtained by the Voyager~1 Faraday cups, however no distributions were obtained.
\citet{tokar00} reported observations of suprathermal FABs upstream of an IP shock observed by the ACE mission \citep[][]{stone98} on 7 April 1998, but the authors could not determine details of the ion distribution functions.

\citet{gosling83} stated that we should not expect to observe non-diffuse ion distributions upstream of IP shocks. The IP shocks have large curvature radii (of the order of 0.5~AU at heliocentric distance of 1~A.U. compared to a few tens of Earth radii, R$_E$, of the Earth's bow-shock) which means that the magnetic field lines stay connected to them for very long times, typically for a day or longer. At planetary shocks these times are of the order of ten minutes. In the case of the planetary shocks we can observe the process of particle acceleration from the beginning, when B-field lines first connect to the bow-shock. In the case of IP shocks we expect to observe acceleration processes at later stages, hence we would detect diffuse ions.
Another problem is that spacecraft are usually not equipped to measure ion distributions continuously from SW thermal to suprathermal energies.

Here we present the first observations of different types of suprathermal ion distributions upstream of a single IP shock that was observed on 8 October 2013, by ACE and the ARTEMIS P1 and P2 spacecraft. We combine the cross-calibrated measurements of the ARTEMIS thermal and energetic particle sensors, obtaining 3D ion distributions covering the key suprathermal energy range. The P1 and P2 spacecraft first observed field-aligned and gyrating ions arriving from the IP shock. As the shock approached, the ion distributions changed to intermediate and then to almost diffuse. These observations confirm that the same ion acceleration mechanisms that are at work at Earth's bow-shock also act at IP shocks. However, in the case of the latter the ions can be accelerated to higher energies compared to those at Earth's bow-shock. 

\section{Datasets}
We use measurements of the two identical ARTEMIS spacecraft orbiting the Moon \citep[][]{angelopoulos10}. Magnetic field measurements are provided by the Fluxgate Magnetometer \citep[FGM, ][]{auster08}. The FGM data are only available in spin (4~s) cadence. Plasma measurements are provided by the Electrostatic Analyzer \citep[ESA, ][]{McFadden08} and Solid State Telescope \citep[SST, ][]{angelopoulos08b}. ESA provides ion measurements between $\sim$5~eV and $\sim$25~keV. SST provides ion data between 25~keV and 6~MeV.

During the time of interest ESA and SST switched from the Fast Survey Mode to the Slow Survey Mode which affects the cadence of the omni-directional ion spectra and of three-dimensional ion distributions. A detailed description of the ESA and SST operational modes and the explanation on how the combined spectra and distributions from both instrument were obtained, are available in the appendix.

We also use the ACE magnetic field data from the MAG instrument \citep{smith98} with 1~second cadence.

All the spacecraft coordinates and measured vectors are given in Geocentric Solar Ecliptic (GSE) coordinate system which is defined so that the X-axis points from the Earth towards the Sun and the Z-axis towards the ecliptic North pole. The Y axis completes the right-hand system.

\section{Observations}
\begin{figure*}
\centering
\includegraphics[width = 0.9\textwidth]{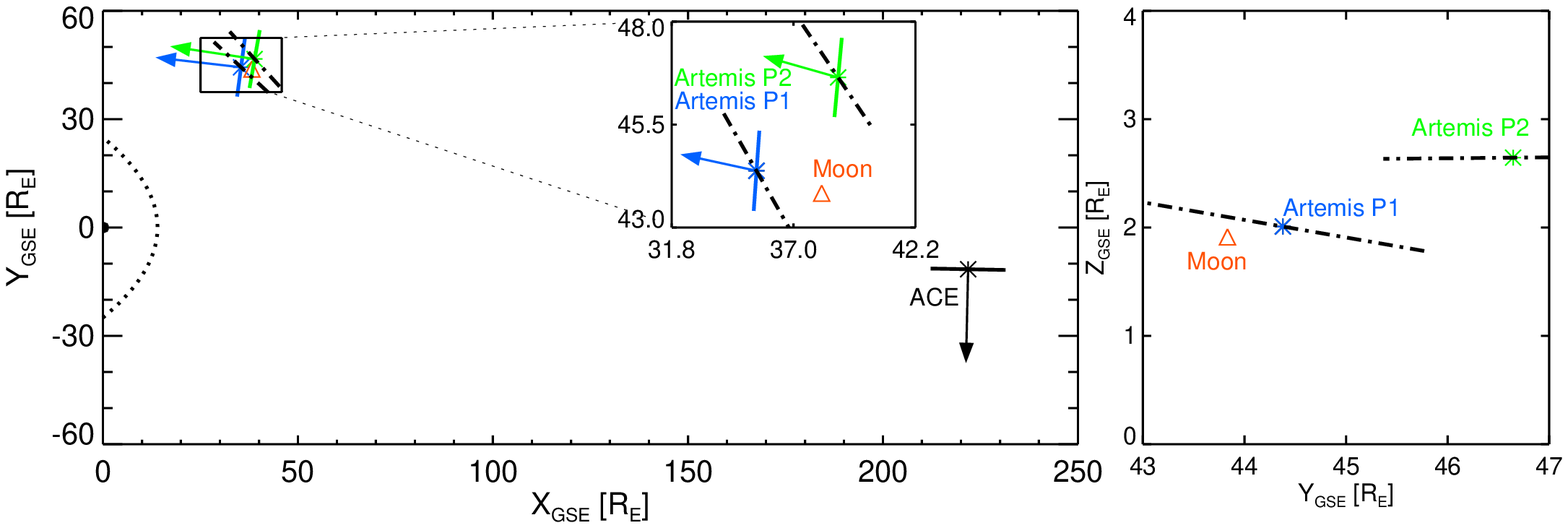}
\caption{Positions of both ARTEMIS spacecraft, the ACE spacecraft and the Moon in GSE XY plane on 8 October 2013 at times of the shock detection. Positions of P1 and P2 spacecraft are marked with blue and green asterisks, respectively. The red triangle marks the position of the Moon. The arrows show the projections of shock normals at each spacecraft. In the case of P1 and P2 the black dash-dotted lines represent the orientation of the IMF. The small black dot at the origin represents the Earth while the black dotted curve surrounding it represents its nominal bow-shock (we use the model bow-shock from \citet{narita04}).}
\label{fig:posxy}
\end{figure*}

We selected the 8 October 2013 IP shock from the {\it Catalog of IP shocks observed in the Earth's neighbourhood by multiple spacecraft between 2011-2014} available at \\
http://usuarios.geofisica.unam.mx/primoz/IPShocks.html. ACE observed the shock at 19:40:49~UT while the ARTEMIS P1 and P2 spacecraft observed it at 20:16:56~UT and 20:16:24~UT, respectively. At the times of the shock passage the three spacecraft were located at: (247.0, -25.0. 0.9)~R$_E$, (56.5, 20.6, -4.6)~R$_E$ and (56.2, 25.7, -4.6)~R$_E$ (Figure~\ref{fig:posxy}).

The separations of the ARTEMIS spacecraft from the Moon were 10.2 and 2.6 lunar radii (R$_L$) along the Sun-Moon line and 2.0~R$_L$ and 10.7~R$_L$ perpendicular to it for P1 and P2, respectively. According to \citet{harada15} these distances are large enough so that no significant Moon-related ion fluxes should be detected by either of the ARTEMIS spacecraft. Also, the IMF orientation indicates that the spacecraft were not magnetically connected to the Moon nor to the Earth's bow-shock (Figure~\ref{fig:posxy}).

The shock normal and the $\theta_{Bn}$ at each spacecraft were calculated using the magnetic coplanarity method \citep[e.g., ][]{schwartz98}: (-0.02, 0.96, -0.27) and 74$^\circ$ at ACE, (-0.81. 0.1, 0.58) and 61$^\circ$ at P1 and (-0.8, 0.13, 0.59) and 47$^\circ$ at P2 (other methods, such as mixed methods \citep{schwartz98} provided very similar results). The $\theta_{Bn}$ values at P1 and P2 do not overlap regardless of the method used. The estimated shock speeds in the spacecraft frame and the Alfv\'enic Mach numbers M$_A$, were calculated to be 428~kms$^{-1}$ and 4.9 at P1 and 456~kms$^{-1}$ and 5.7 at P2. The $\theta_{Bn}$ was smaller at P2, where the M$_A$ was higher. While the shock normal directions are similar at P1 and P2, at ACE the normal differs by 90$^\circ$. This is not surprising since it was shown by \citet{szabo05} that the IP shock normals may differ greately when the spacecraft separations perpendicular to the Sun-Earth line are of several tens of R$_E$.

\subsection{Reflected ions}
\begin{figure*}
\centering
\includegraphics[width = 0.7\textwidth]{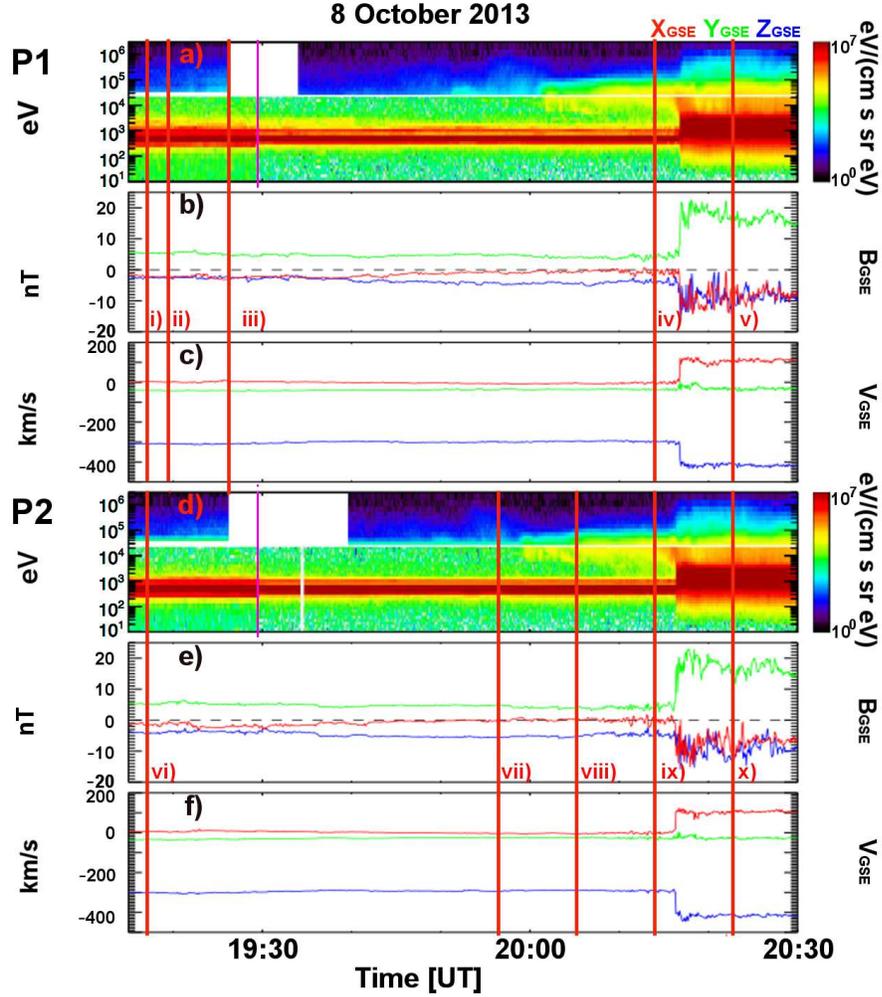}
\caption{ARTEMIS P1 and P2 observations. a) and d): dynamic ion spectra from SST and ESA. b) and e): interplanetary magnetic field components. c) and f): SW velocity components. The red, green and blue curves on IMF and velocity panels represent the X$_{GSE}$, Y$_{GSE}$ and Z$_{GSE}$ components, respectively. Roman numbers and vertical red lines mark the times of ion distributions shown in Figure~\ref{fig:dist}. Vertical purple lines mark the times when the ion instruments switched the operational modes. 
}
\label{fig:timeserie}
\end{figure*}

\begin{figure*}
\centering
\includegraphics[width = 1.0\textwidth]{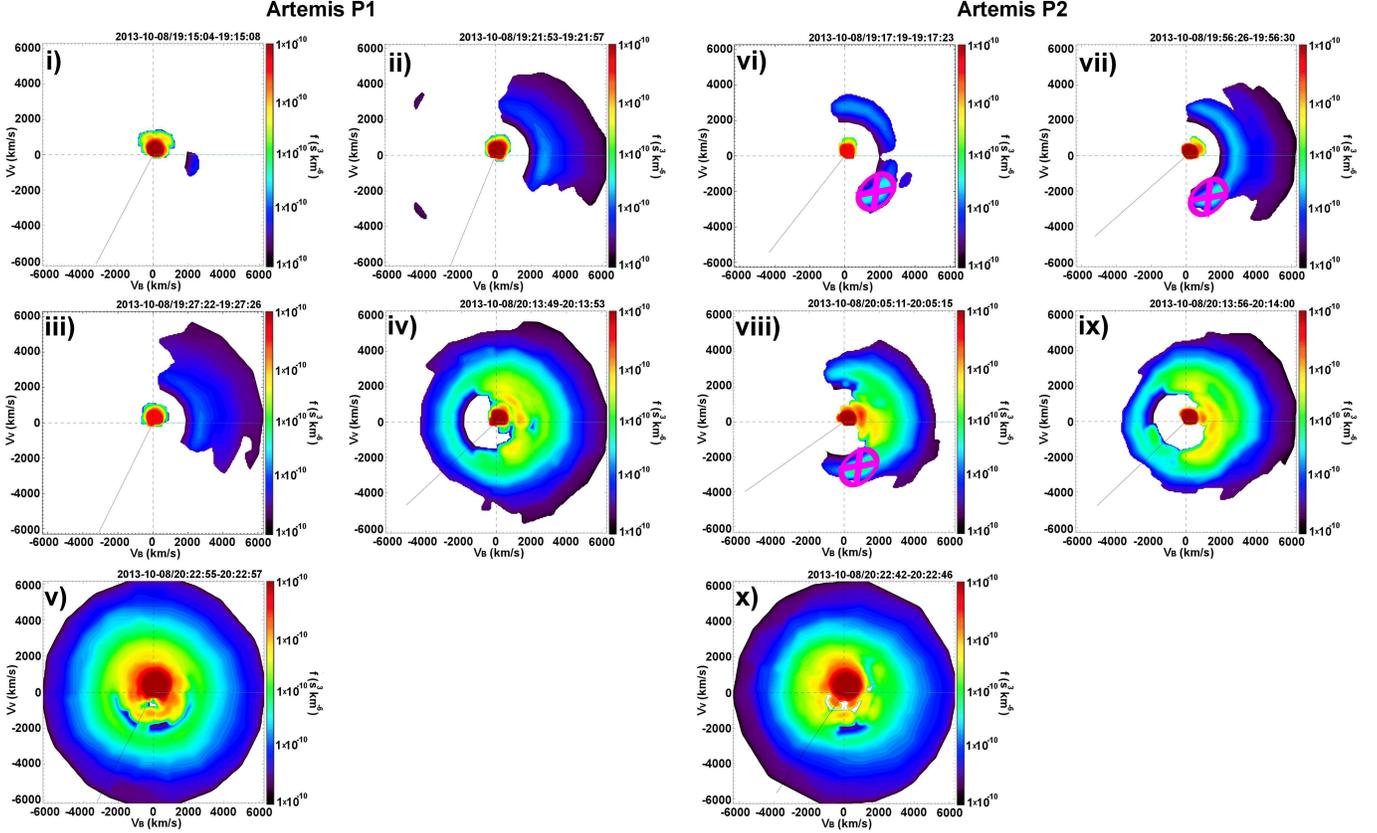}
\caption{Ion distribution functions measured by ARTEMIS P1 (i-v) and P2 (vi-x) spacecraft at five different times. The distribution slice planes are defined so that they contain the IMF and the SW velocity vectors. The x-axis (V$_B$) points along the IMF and the y-axis (V$_V$) points along the SW velocity component perpendicular to the IMF. The black lines show the Sun direction. Colors represent the logarithm of the particle phase space density. Crossed purple ellipses in panels vi), vii) and viii) mark the signal due to photons reflected from the lunar surface.}
\label{fig:dist}
\end{figure*}

The 8 October 2013 shock was driven by a complex event composed of a SIR and at least one ICME. Figure~\ref{fig:timeserie} shows ARTEMIS P1 (panels a - c) and P2 (panels d - f) observations from 19:10~UT to 20:30~UT. The panels a) and d) exhibit combined SST and ESA ion spectra (the colors represent the logarithm of the particle energy flux), panels b) and e) exhibit IMF components and c) and f) panels show the SW velocity components. The red vertical lines and roman numerals show times of the distributions exhibited in Figure~\ref{fig:dist}.

Figure~\ref{fig:dist} shows particle (ion) distribution functions (PDF) at five different times obtained by P1 (panels i - v) and P2 (panels vi - x) spacecraft. In both cases there are four PDFs observed upstream and one downstream of the shock. Note that the ion spectra in Figure~\ref{fig:timeserie} and PDFs in Figure~\ref{fig:dist} were made with different datasets resulting in some discrepancies between the two figures (see appendix).

On panels a) and d) of Figure~\ref{fig:timeserie} we can see a red trace centered at $\sim$470~eV, which is the SW. It corresponds to the red circular spot on all panels in Figure~\ref{fig:dist}.
The FABs are barely detected by ESA, but they appear as a light-blue trace at $\lesssim$200~keV in the SST part of the spectra.

In all panels of Figure~\ref{fig:dist}, part of the ion PDF around the SW core is missing. This occurs when the intensity of suprathermal ions is less than the sensitivity of the instrument. Figure~\ref{fig:dist1D} in the appendix illustrates this by showing the signal from both instruments and their corresponding one-count levels. For ESA, the intensity of the reflected suprathermal ions was mostly below the one-count level, except during the last $\sim$15 minutes before the shock crossing. In contrast, the lowest energy channels of SST are much more sensitive and can detect these suprathermal ions.

We first look at the P1 distributions and ion spectra. Figure~\ref{fig:dist}i) shows the first particle distribution function featuring FABs during the time interval centered at 19:15:06~UT. The FABs appear as a blue and purple trace with velocities at V$_B\sim$2000~kms$^{-1}$ and V$_V$ between -600~kms$^{-1}$ and 100~kms$^{-1}$. These velocities correspond to energies of $\sim$21~keV in the spacecraft frame. We also calculate suprathermal ion kinetic energies in the shock rest frame by substracting the shock velocity with respect to the spacecraft (428~kms$^{-1}$ along the shock normal) but the result remains roughly the same. Such kinetic energies of the FABs are much higher than in the case of the Earth's bow-shock, where the FABs exhibit energies less than 10~keV \citep{thomsen85}. It seems that although the IP shock studied here had a lower M$_A$ than the typical Earth's bow-shock near its subsolar point, the IP shock is able to accelerate the FABs to much higher energies. This is probably related to IP shock's large curvature radii and long connection times of the IMF field lines to the shock. Ions that reflect at quasi-perpendicular section of the IP shock remain at such section for longer periods and consequently the shock drift and shock surfing acceleration mechanisms act for longer periods accelerating ions to higher energies.

The flux of the reflected ions in Figure~\ref{fig:timeserie}a) intensifies with time and their maximum energy increases and eventually reaches $\sim$200~keV. Figure~\ref{fig:dist}ii) shows the PDF  at 19:21:55~UT. We can see that the ion beam has broadened. The peak of the distribution lies along the magnetic field but the beam extends to the upper right quadrant. The maximum velocities of the ions are $\sim$6000~kms$^{-1}$ in the spacecraft frame, corresponding to energies of $\sim$190~keV. Just before the shock (panel iv) the ion PDF becomes diffuse, as revealed by the SST measurements.
Downstream of the shock (panel v) the ions are heated and their distributions become isotropic. 

In the case of the P2 spacecraft the observed PDFs look a bit different. First, we note an intense spot in lower-right quadrant on panels vi)-viii) marked by a crossed purple ellipse. A careful inspection of the PDFs in the X$_{GSE}$-Y$_{GSE}$ plane revealed that this signal comes from the direction of the Moon. It is not related to any ions but it is caused by the reflected photons coming from the Moon, so we will disregard it. We still see ions in the upper right cuadrant on panel vi). These are non-gyrotropic ions. At later times (panels vii and viii) we observe intermediate ion PDFs and just before the shock arrival (panel ix) the ion PDF is almost completely diffuse. Again, downstream of the shock we observe an isotropic, heated ion PDFs (panel x).

\subsection{Upstream waves}
\begin{figure*}
\centering
\includegraphics[width = 0.5\textwidth]{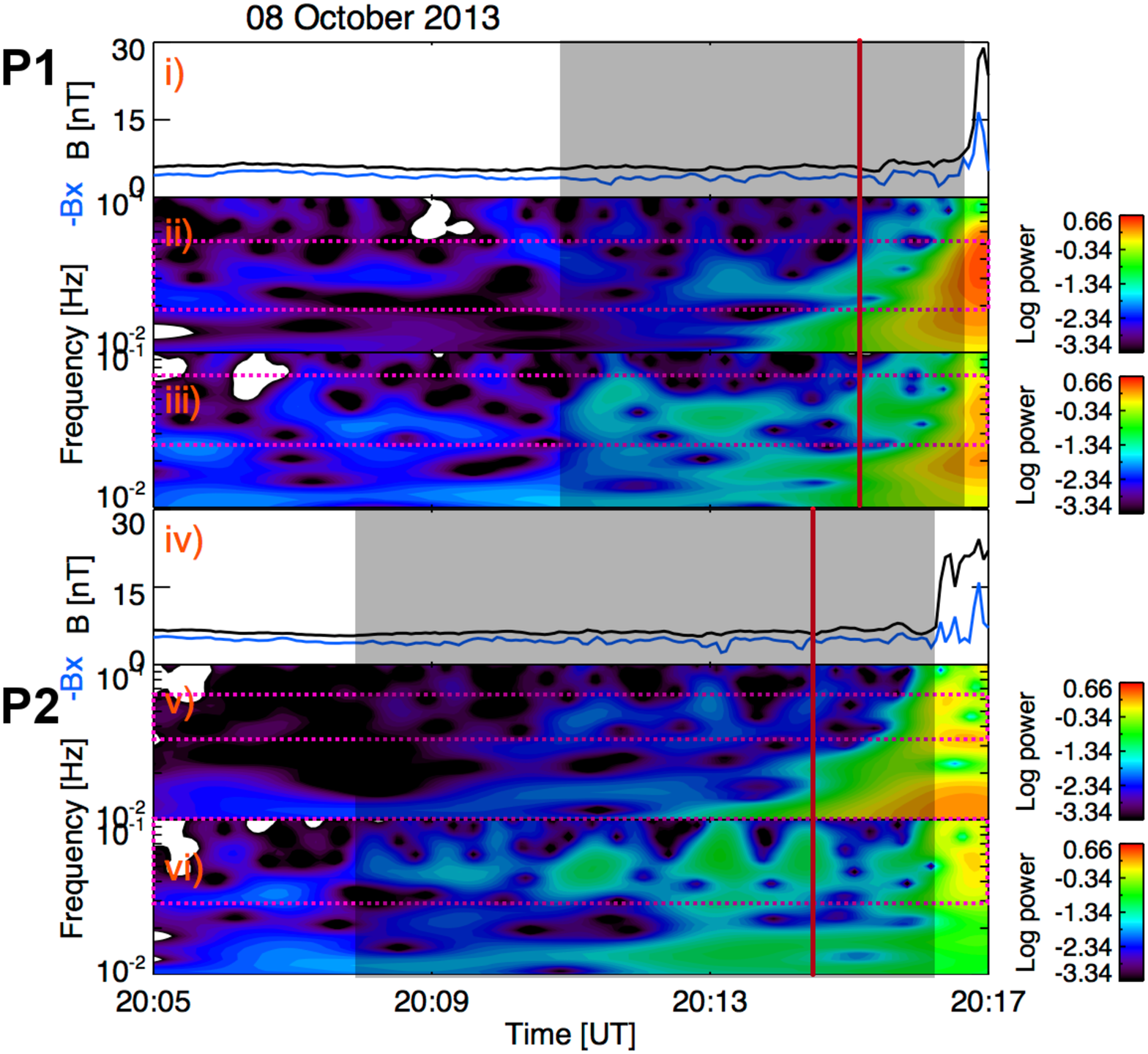}
\caption{i) and iv): IMF magnitude (black) and -Bx component (blue) from ARTEMIS P1 and P2 spacecraft, respectively. ii) and v): Wavelet spectra of the IMF magnitude. iii) and vi): Wavelet spectra of the Bx component. Horizontal purple lines delimit the frequency range of interest. Vertical red lines show times of diffuse ions from Figure~\ref{fig:dist}. The intervals of upstream ULF waves are shaded in gray.}
\label{fig:wavelet}
\end{figure*}

Figure~\ref{fig:wavelet} shows B magnitude (black) and -B$_{x,GSE}$ component (blue) on panels i) and iv) (corresponding to P1 and P2 observations, respectively). Panels ii) and v) show wavelet spectra of B magnitude while panels iii) and vi) show the spectra for the B$_{x, GSE}$ component. The shaded intervals correspond to times when the upstream ULF waves are present. The waves appear $\sim$7.8~minutes before the shock arrival in the case of P2 and $\sim$5.2~minutes before in the case of P1. At first they are highly transverse, but they become more compressive closer to the shock front. Their frequencies are between 0.02~Hz and 0.1~Hz (periods between 10~s and 50~s). By comparing Figures~\ref{fig:timeserie} and \ref{fig:wavelet} we can see that the FABs coincide with times when no ULF waves are present, but that the almost diffuse ion PDFs appear together with upstream ULF waves that exhibit an important compressive component.

\section{Discussion and conclusions}
We report the first observations of different suprathermal ion distributions upstream of the single 8 October 2013 IP shock. These observations were made with the two ARTEMIS spacecraft. The shock properties, the ion PDFs and the upstream ULF wave foreshocks differ at the two observational points. The shock is weaker and quasi-perpendicular (M$_A$=4.9, $\theta_{Bn}$=61$^\circ$) at P1, while it is stronger and less quasi-perpendicular (M$_A$=5.7, $\theta_{Bn}$=47$^\circ$) at P2. Consequently, at P2 the ULF waves appear before than at P1 and they are more intense.

Ion distributions vary from FABs (at P1) and gyrating ions (at P2) upstream of the shock, to intermediate and finally to diffuse distributions just before the shock arrival. The FABs and the gyrating ions are observed in the absence of any ULF fluctuations, while the diffuse ions coincide with partially compressive ULF waves.

The energies of the FABs in the shock rest frame are of the order of 20~keV, which is much more than in the case of the Earth's bow-shock, where they are $\lesssim$10~keV. This is probably a consequence of larger curvature radii of IP shocks and longer connection times of IMF lines to the IP shock surface. Under these conditions ions travel larger distances with $\theta_{Bn}<60^\circ$ meaning that the shock drift and shock surfing mechanisms \citep[][]{hudson65, lever01} accelerate them to higher energies.

In addition to the curvature radius and M$_A$, there are other factors that influence the efficiency of ion acceleration at shocks, such as background turbulence and plasma beta. One should also keep in mind that the quasi-perpendicular, supercritical shocks undergo continuous self-reformation  and this shock nonstationarity additionally impacts the ion reflection and energization \citep[e.g., ][]{mazelle10, lobzin07, yang09}.

The energies of the observed diffuse ions are $\lesssim$200~keV, which is similar to ions near the Earth's bow-shock.

\acknowledgments
This work has been supported by the International Space Science Institute (ISSI). PK's and XBC's work was supported by DGAPA/PAPIIT grants IA104416 and IN105014 and by CONACYT grant 179588. HH's work was supported by NASA contract NAS5-02099.

\appendix
\section{Contents}
This section contains information on how the combined ion omni directional spectra and three dimensional particle distribution functions were obtained from the data from the ESA and SST instruments and an explanation on their operational modes.
We also show the sensitivities (one count levels) of both instruments and compare them with the observations.

\section{ESA and SST operational modes}
The ARTEMIS ESA and SST instruments were in magnetospheric Fast Survey Mode until $\sim$19:29~UT. After that they were in magnetospheric Slow Survey Mode. While each ESA and SST sample is always collected over one spacecraft spin period ($\sim$4 seconds), during the two modes, there are differences in the angular, energy, and temporal resolutions of various downlinked data products. 

During Fast Survey, we have three-dimensional ESA ``full mode'' ion distributions (88 angles, 32 energies) available every 32 spins ($\sim$2.1 minutes) and ``reduced mode'' ion distributions (50 angles, 24 energies) available for every spin. SST ``full mode'' ion distributions (64 angles, 16 energies) are also available for every spin.

During Slow Survey mode, ESA ``full mode'' ion distributions are available every 128 spins ($\sim$9 minutes), and ``reduced/omni-directional'' distributions (1 angle, 32 energies) are available every spin. SST ``full mode'' ion distributions are available every 64 spins ($\sim$4.3 minutes) and ``reduced/omni-directional'' distributions (1 angle, 16 energies) are available for every spin.

The mode change may sometimes result in a minor data loss of some products. Note also that some ground calibrations are only possible for the higher angular resolution data products.

\section{Combined omni-directional ion spectra and three-dimensional ion distributions}

The high time-resolution omni-directional ion spectra shown in Figure 2 of this study were constructed in the following way. During the early part of the interval, when the spacecraft were in Fast Survey mode, we have plotted the spectra of ESA reduced mode distributions and SST full mode distributions in the same panel (no interpolation, as evidenced by the small white horizontal gap). During the later part of the interval, when the spacecraft were in Slow Survey mode, we have plotted the ESA and SST reduced/omni-directional distributions in the same panel (no interpolation). There were minor losses of the reduced/omni-directional data products during the mode change, which manifest as the white gaps in Figure 2 panels a and d.

To make the ion distribution slices shown in Figure 3, we have combined the ESA and SST full mode (highest energy and angular resolution) measurements using 3D interpolation. (Note that the cadence at which these measurements are available depends on which Survey mode the instruments were in, as described above.) This type of combined distributions have recently been used in several ARTEMIS/THEMIS studies of different plasma regions \citep[see e.g. ][]{dorfman17, hietala15, hietala17, runov15}. We first removed the bins that were at or below the one-count-level from the measurements. We then combined the (cleaned-up) ESA and SST measurements by interpolating in 3D across the energy gap (at 25~keV) between the instruments. The lowest SST energy channels ($<$35~keV) on P2 were excluded from the interpolation due to degradation effects. Note that the distribution slices only show the features that are still above the one-count-level after the observed (and cleaned-up) distribution has been interpolated into the slice plane. 

\section{Sensitivities of the ESA and SST instruments}

\begin{figure*}[h!]
\centering
\includegraphics[angle=0, width = 1.0\textwidth]{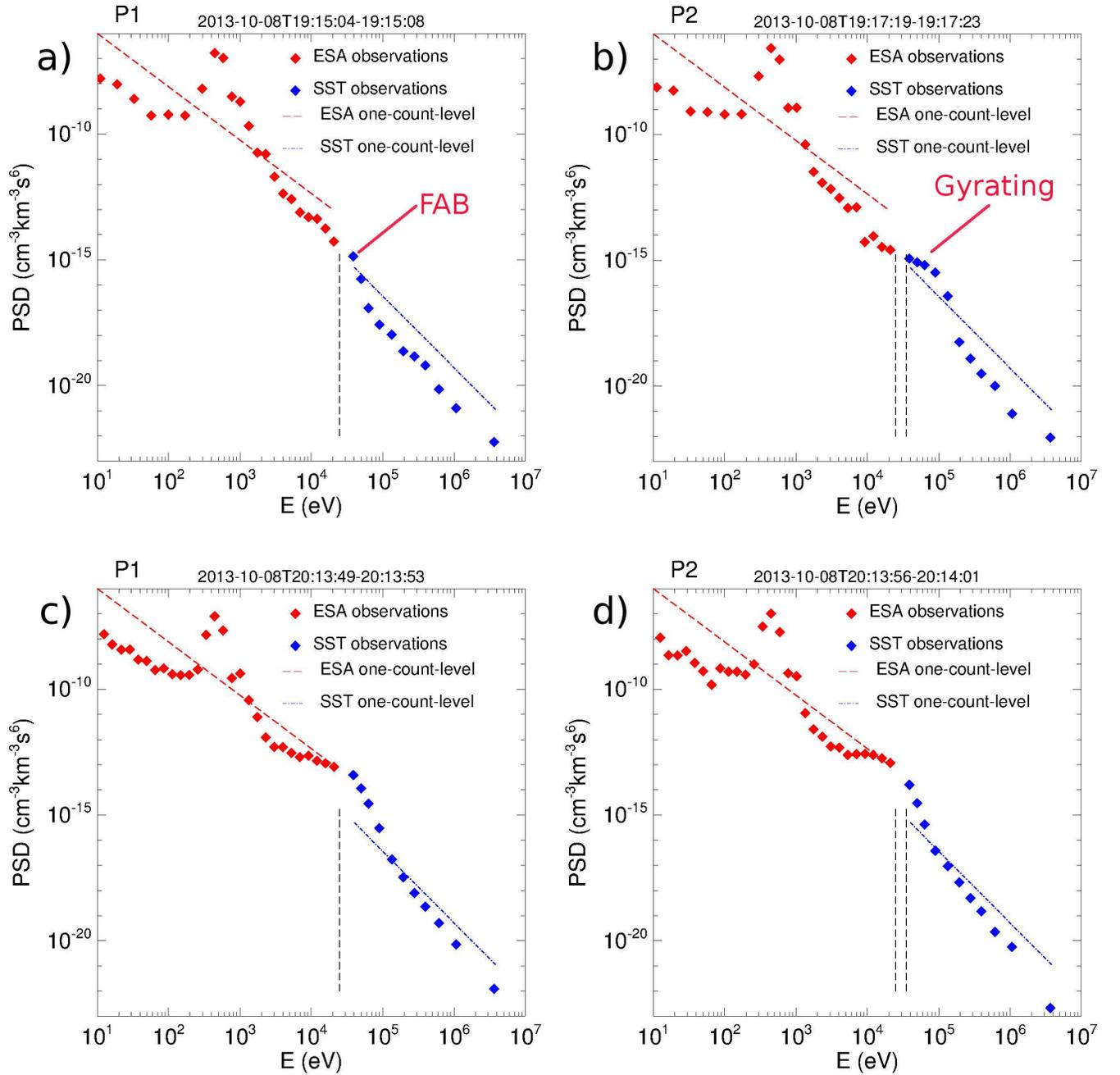}
\caption{1D spectra obtained by ARTEMIS P1 spacecraft at 19:15:04-19:15:08~UT (a) and 20:13:49-20:13:53~UT (c) and by P2 spacecraft at 19:17:19-19:17:23~UT (b) and 20:13:56-20:14:01~UT (d) on 8 October 2013. Red and blue diamonds represent ESA and SST data, respectively. The red dashed lines represent the one-count level of the ESA instrument and the blue dash-dotted lines show the one-count level of the SST instrument. Vertical dashed black lines delimit ESA data from SST data. FAB and gyrating ions are marked on panels (a) and (b).}
\label{fig:dist1D}
\end{figure*}

Figure~\ref{fig:dist1D} shows four one-dimensional spectra from ARTEMIS P1 spacecraft obtained at 19:15:04-19:15:08~UT (a) and 20:13:49-20:13:53~UT (b) and from P2 spacecraft at times 19:17:19-19:17:23~UT (b) and 20:13:56-20:14:01~UT (d) on 8 October 2013. The red and blue diamonds show measurements of the ESA and SST instruments, respectively. The red dashed lines and blue dash-dotted lines represent one-count levels of the ESA and SST, respectively. We mark FAB and gyrating ions on panels (a) and (b).

We see that, on average, the sensitivity of the ESA instruments does not permit the detection of the suprathermal ions with energies between 2~keV and 20~keV for the IP shock studied here. Similarly, the SST instrument does not observe ions with energies above 200~keV.

\begin{thebibliography}{99}
\bibitem[Angelopoulos(2008a)]{angelopoulos08a} Angelopoulos, V., Sibeck, D. G., Farrell, W. M. et al.\ 2008, NLSI Lunar Science Conference, July 20-23, LPI
Contribution No. 1415, abstract no. 2161.
\bibitem[Angelopoulos(2008b)]{angelopoulos08b} Angelopoulos, V., Sibeck, D. G., Carlson, C. W. et al.\ 2008, \ssrv, 141, 453, doi:10.1007/s11214-008-9378-4.
\bibitem[Angelopoulos(2010)]{angelopoulos10} Angelopoulos, V.\ 2010, \ssrv, 165, 3, doi:10.1007/s11214-010-9687-2.
\bibitem[Auster et al.(2008)]{auster08} Auster, H.~U., Glassmeier, K.~H., Magnes, W. et al.\ 2008, \ssrv, 141, 235, doi:10.1007/s11214-008-9365-9.
\bibitem[Armstrong et al.(1970)]{armstrong70} Armstrong, T.~P., Krimigis,  S.~M., Behannon K.~W.\ 1970, \jgr, 75(31), 5980, doi:10.1029/JA075i031p05980.
\bibitem[Bagenal(1992)]{bagenal92} Bagenal F.\ 1992, Annu. Rev. Earth Planet Sci., 20, 289
\bibitem[Bavassano-Cattaneo et al.(1986)]{bavassano86} Bavassano-Cattaneo, M.~B., Tsurutani, B.~T., Smith, E.~J., Lin, R.~P.\ 1986, \jgr, 91(A11), 11929, doi:10.1029/JA091iA11p11929.
\bibitem[Cohen(2006)]{cohen06} Cohen, C. M. S.\ 2006, in Solar Eruptions and Energetic Particles, Vol. 165 ed. N. Gopalswamy, R. Mewaldt, \& J. Tortsi, (Washington, D.C.: AGU), 275
\bibitem[Cravens and Gombosi(2004)]{cravens04} Cravens, T.~E, Gombosi, T.~I.\ 2004, \adspr, 33, 196, doi:10.1016/j.asr.2003.07.053.
\bibitem[Dorfman et al.(2017)]{dorfman17} Dorfman, S., Hietala, H., Astfalk, P. and Angelopoulos, V.\ 2017, \grl, 44, 2120, doi:10.1002/2017GL072692.
\bibitem[Eastman et al.(1981)]{eastman81} Eastman, T.~E. Anderson, R.~R., Frank, L.~A.,Parks,  G. K.\ 1982,\jgr, 86, 4379.
\bibitem[Eastwood et al.(2005)]{eastwood05} Eastwood, J.P., Lucek, E.A., Mazelle, C. et al.\ 2005, \ssrv, 118, 41, doi:10.1007/s11214-005-3824-3
\bibitem[Edminston and Kennel(1986)]{edminston86} Edmiston, J.~P., Kennel, C.~P.\ 1986, \jgr, 91(A2), 1361 doi:10.1029/JA091iA02p01361.
\bibitem[Fuselier al.(1986)]{fuselier86} Fuselier, S.~A., Thomsen, M.~F., Gosling, J.~T., Bame, S.~J.\ 1986, \jgr, 91, 91.
\bibitem[Gosling et al.(1978)]{gosling78}  Gosling, J.~T., Asbridge, J.~R., Bame, S.~J., Paschmann, G., Sckopke, N.\ 1978, \grl, 5, 957, doi:10.1029/GL005i011p00957.
\bibitem[Gosling et al.(1979)]{gosling79} Gosling, J.~T., Asbridge,  J.~R., Bame,  S.~J., Paschmann, G., Sckopke, N.\ 1978, \grl, 5,957.
\bibitem[Gosling(1983)]{gosling83} Gosling, J.~T.\ 1983, \ssrv, 34, 113, doi:10.1007/BF00194621.
\bibitem[Gosling et al.(1984)]{gosling84} Gosling, J.~T., Bame,  S.~J., Feldman, W.~C.,Paschmann, G., Sckopke, N., Russell, C.~T.\ 1984, \jgr, 89(A7), 5409, doi:10.1029/JA089iA07p05409.
\bibitem[Gosling and Robson(1985)]{gosling85} Gosling, J.~T., Robson, A.~E.\ 1985, Ion Reflection, Gyration, and Dissipation at Supercritical Shocks, in Collisionless Shocks in the Heliosphere: Reviews of Current Research (eds B. T. Tsurutani and R. G. Stone), American Geophysical Union, Washington, D. C.. doi: 10.1029/GM035p0141
\bibitem[Gosling and Pizzo(1999)]{gosling99} Gosling, J.~T., Pizzo, V.~J.\ 1999, \ssrv, 89, 21, doi:10.1023/A:1005291711900.
\bibitem[Gurgiolo et al.(1981)]{gurgiolo81} Gurgiolo, C., Parks, G.~K., Mauk, B.~H., Lin,  C.~S., Anderson,  K.~A., Lin,  R.~P., Reme, H.\ 1981, \jgr, 86, 4415.
\bibitem[Gurgiolo et al.(1983)]{gurgiolo83} Gurgiolo, C., Parks, G.~K., Mauk, B.~H.\ 1983, \jgr, 88, 9093.
\bibitem[Harada et al.(2015)]{harada15} Harada, Y., Halekas, J.~S., Poppe, A.~R., Tsugawa, Y., Kurita, S., McFadden, J.~P.\ 2015, \jgr, 120, 4907, doi:10.1002/2015JA021211.
\bibitem[Hietala et al.(2015)]{hietala15} Hietala, H., Drake,  J. F., Phan, T. D., Eastwood, J. P. and McFadden, J. P.\ 2015\grl, 42, 7239, doi:10.1002/2015GL065168.
\bibitem[Hietala et al.(2017)]{hietala17} Hietala, H., Artemyev, A. V. and Angelopoulos, V.\ 2017, \jgr, 122, 2010, doi:10.1002/2015JA021166.
\bibitem[Hudson and Kahn(1965)]{hudson65} Hudson, P. D. and Kahn, F. D.\ 1965, \mnras, 131, 23-49.
\bibitem[Kahler(2003)]{kahler03} S.W. Kahler, S.~W.\ 2003, \adspr, 32, 2587, doi:10.1016/j.asr.2003.02.006.
\bibitem[Kis et al.(2004)]{kis04} Kis, A., Scholer, M., Klecker, B., M\"obius, E, Lucek, E. A., R\`eme, H., Bosqued, J. M., Kistler, L. M. and Kucharek, H.\ 2004, \grl, 31, L20801, doi:10.1029/2004GL020759.
\bibitem[Kis et al.(2007)]{kis07} Kis, A., Scholer, M., Klecker, B., Kucharek, H., Lucek, E. A., and R\`eme, H.\ 2007, \angeo, 25, 785, doi:10.5194/angeo-25-785-2007.
\bibitem[Lever et al.(2001)]{lever01} Lever, E. L., Quest, K. B., Shapiro, V. D.\ 2001, \grl, 28: 1367m, doi:10.1029/2000GL012516.
\bibitem[Lobzin et al.(2007)]{lobzin07} Lobzin, V. V., Krasnoselskikh, V. V., Bosqued, J.-M., et al.\ 2007, \grl, 34, L05107
\bibitem[Luhmann et al.(2004)]{luhmann04} Luhmann, J.~G., Ledvina, S.~A., Russell, C.~T.\ 2004, \adspr, 33, 1905, ISSN 0273-1177, doi:10.1016/j.asr.2003.03.031.
\bibitem[Manchester et al.(2005)]{manchester05} Manchester W. B. IV, Gombosi T. I., Zeeuw D. L. et al.\ 2005, \apj, 622, 1225
\bibitem[Mazelle et al.(2010)]{mazelle10}Mazelle, C., Lembege, B., Morgenthaler, A., et al.\ 2010, in AIP Conf. Proc. 1216, 12th International Solar Wind Conference (Melville, NY: AIP), 471
\bibitem[McFadden et al.(2008)]{McFadden08} McFadden, J.~P., Carlson, C.~W., Larson, D. et al.\ 2008, \ssrv, 141, 277. doi:10.1007/s11214-008-9440-2.
\bibitem[Meziane et al.(2013)]{meziane13} Meziane, K., Hamza, A. M., Wilber, M., Mazelle,  C. and Lee, M. A.\ 2013, \jgr, 118, 6946, doi:10.1002/2013JA019060.
\bibitem[Narita et al.(2004)]{narita04} Narita, Y., Glassmeier, K.-H., Schafer, S., et al.\ 2004, \angeo, 22, 2315, doi:10.5194/angeo-22-2315-2004.
\bibitem[Paschmann et al.(1979)]{paschmann79} Paschmann, G., Sckopke, N., Bame, S.~J., Gosling, J.~T., Russell, C.~T., Greenstadt, E.~W.\ 1979, \grl, 6,209.
\bibitem[Paschmann et al.(1982)]{paschmann82} Paschmann, G., Sckopke, N., Bame, S.~J., Gosling, J.~T.\ 1982, \grl, 9, 881.
\bibitem[Reames et al.(1996)]{reames96} Reames, D.~V., Barbier, L.~M., Ng, C.~K.\ 1996, \apj, 466, 473, doi:10.1086/177525.
\bibitem[Runov et al.(2015)]{runov15} Runov, A., Angelopoulos, V., Gabrielse, C., Liu, J., Turner, D. L. and Zhou, X.-Z.\ 2015, \jgr, 120, 4369, doi: 10.1002/2015JA021166.
\bibitem[Russell(1993)]{russell93} Russell, C.~T.\ 1993, \jgr, , 98(E10), 18681, doi:10.1029/93JE00981.
\bibitem[Savoini et al.(2013)]{savoini13} Savoini, P., Lembege, B., Stienlet, J.\ 2013, \jgr, 118, 1132, doi:10.1002/jgra.50158.
\bibitem[Schwartz(1998)]{schwartz98} Schwartz, S. J.\ 1998, in ESA/ISSI 1, 249
\bibitem[Schwenn(2006)]{schwenn06} Schwenn, R.\ 2006, LRSP, 3, 2
\bibitem[Sheeley et al.(1985)]{sheeley85} Sheeley Jr., N. R., R. A. Howard, M. J. Koomen, D. J. Michels, R. Schwenn, K. H. Mühlhäuser, Rosenbauer, H.\ 1985, \jgr, 90(A1), 163, doi:10.1029/JA090iA01p00163.
\bibitem[Smith et a.(1998)]{smith98} Smith, C.~W., Acu\~na, M.~H., Burlaga, L.~F., L'Heureux, J., Ness, N.~F., Scheifele, J.\ 1997, 86, 613, doi:10.1023/A:1005092216668.
\bibitem[Stone et al.(1998)]{stone98} Stone, E.~C., Frandsen, A.~M., Mewaldt, R.~A., Christian, E.~R., Margolies, D., Ormes, J.~F., Snow, F.\ 1998, \ssrv, 86, 1, doi:10.1023/A:1005082526237.
\bibitem[Szabo(2005)]{szabo05} Szabo, A.\ 2005, AIP Conference Proceedings 781, 37, doi: http://dx.doi.org/10.1063/1.2032672.
\bibitem[Thomsen(1985)]{thomsen85} Thomsen, M.~F.\ 1985,. Upstream Suprathermal Ions, in Collisionless Shocks in the Heliosphere: Reviews of Current Research (eds B. T. Tsurutani and R. G. Stone), American Geophysical Union, Washington, D. C.. doi: 10.1029/GM035p0253
\bibitem[Tokar et al.(2000)]{tokar00} Tokar, R.~L., Gary, S.~P., Gosling, J.~T., McComas, D.~J., Skoug, R.~M., Smith, C.~W., Ness, N.~F., Haggerty, D.\ 2000, \jgr, 105, 7521
\bibitem[Vi\~nas et al.(1984)]{vinas84} Vi\~nas, A.~F., Goldstein, M.~L., Acu\~na, M.~H.\ 1984, \jgr, 89, 3762
\bibitem[Williams et al.(1998)]{williams98} Williams, D., Leske, R., Mewaldt, R. et al.\ 1998, \ssrv, 85, 379, doi:10.1023/A:1005143710585.
\bibitem[Winske et al.(1984)]{winske84} Winske, D., Wu, C.~S., Li, Y.~Y., Zhou G.~S.\ 1984 \jgr, 89, 7327.
\bibitem[Yang et al.(2009)]{yang09} Yang, Z. W., Lu, Q. M., Lembege, B., Wang, S.\ 2009, \jgr, 114, A03111
\end{thebibliography}
\end{document}